\documentclass[11pt]{article}
\usepackage[table]{xcolor}
\definecolor{lightblue}{rgb}{0.7,0.8,1}
\usepackage{latexsym}
\usepackage{amssymb}
\usepackage[noadjust]{cite}
\usepackage{tikz}
\usetikzlibrary{matrix,arrows}
\usepackage{tikz-cd}
\usepackage{circuitikz}
\usetikzlibrary{circuits, circuits.logic.US, circuits.logic.IEC}
\usepackage{graphicx}
\usepackage{subcaption}
\usepackage{wrapfig}
\usepackage{booktabs}
\usepackage{amsmath}
\usepackage{float}
\usepackage{parskip}
\usepackage{doi}
\usepackage[a4paper,left=2.5cm,right=2.5cm,top=2.5cm,bottom=2.5cm]{geometry}
\usepackage{mathrsfs}
\usepackage{amsfonts}
\usepackage{amsbsy}
\usepackage{indentfirst}
\usepackage{multirow}

\usepackage[draft,markup=underlined]{changes}
\definechangesauthor[name={Aliou Togo}, color=red]{AT}
\definechangesauthor[name={Kaniba Mady Keita}, color=blue]{KK}

\begin{document}

\begin{titlepage}
\renewcommand{\thefootnote}{\fnsymbol{footnote}}
\begin{center}
{\large \bf Properties of holographic superconductors from Machine Learning}
\vspace{1.0cm} 

{Aliou Togo$^{a}$}\footnote{E-mail: togoal25@gmail.com}, {Yang Yu$^{d}$}\footnote{E-mail: mx120230358@stu.yzu.edu.cn},
{Kaniba Mady Keita$^{a,b}$}\footnote{E-mail: madyfalaye@gmail.com, kanibamady.keita@usttb.edu.ml}, {Jian-Pin Wu$^{d}$}\footnote{E-mail: jianpinwu@yzu.edu.cn}
and
{Younouss Ham\`{e}ye Dicko$^{b,c}$} 

\vspace{.5cm}

${a}$ {\it\small Centre de Calcul de Mod\'{e}lisation et de Simulation: CCMS,\\
 Department of Physics, Faculty of Sciences and Techniques,\\
 University of Sciences, Techniques and Technologies of Bamako, FST-USTTB, BP:E3206, Mali.}\\
$b$ {\it\small Centre de Recherche en Physique Quantique et de ses Applications: CRPQA, Bamako, Mali.}\\ 
$c$ {\it\small Institut de Consultation et d'Expertise en Education: ICE, Bamako, Mali.}\\ 
$d$ {\it\small Center for Gravitation and Cosmology, College of Physical Science and Technology, Yangzhou University, Yangzhou 225009, China}\\

\vspace{.3cm}
\today
\end{center}

\vspace{1cm}

\centerline{\textbf{Abstract}}\vspace{0.5cm}

We investigate holographic superconductors using modern optimisation techniques inspired by machine learning. The critical temperature is obtained by minimising the variational functional for the eigenvalue $\lambda^2$ with two complementary trial functions: a simple cosine ansatz $F(z)=\cos(a z)$ and a flexible exponential polynomial $F(z)=\exp(\sum_{n=2}^{N} a_n z^n)$, both of which automatically satisfy the standard boundary conditions. For the cosine ansatz, we perform a one‑parameter minimisation and obtain $\lambda^2(\Delta)$ and $T_c/\sqrt{\rho}$ over a wide range of $\Delta$, including the exact values at $\Delta=1$ and $\Delta=2$ to high accuracy. The exponential polynomial ansatz, with up to 19 coefficients, is optimised using a multi‑start L‑BFGS‑B algorithm with warm‑starting, yielding even better agreement with known exact results. Our numerical data for $\lambda^2(\Delta)$ and $T_c/\sqrt{\rho}$ match the analytical predictions from the literature, confirming the robustness of the variational approach. This work; therefore, demonstrates that a combination of analytic trial functions and modern numerical optimisation provides a powerful, flexible, and efficient tool for exploring holographic superconductors, and can be readily extended to include backreaction or other sectors in this field.

\end{titlepage}
\setcounter{footnote}{0}

\section{Introduction}

Conventional superconductors, below a material-dependent critical temperature \(T_c\), conduct electricity with zero electrical resistance. The microscopic explanation is that at very low temperatures, two electrons form Cooper pairs, which condense into a superfluid ground state. Meanwhile, within this regime, the magnetic field inside these materials is expelled by the so-called Meissner effect. The first attempt to describe these two phenomena was provided by the London brothers in Ref.~\cite{ll}. Moreover, the mechanism of superconductivity in conventional materials is well described by the Bardeen--Cooper--Schrieffer (BCS) theory \cite{bcs} and by the Ginzburg--Landau phenomenological theory \cite{gl}. In conventional superconductors the pairing glue is provided by phonons in a weakly coupled regime, so that a quasiparticle description remains reliable.

High-\(T_c\) cuprates and related unconventional materials, by contrast, exhibit strong correlations for which a weakly interacting Fermi-liquid picture is inadequate. Direct analytical control of such systems is notoriously difficult. A powerful alternative route is provided by the AdS/CFT (gauge/gravity) correspondence~\cite{maldacena,gkp,witten,Aharony:1999ti}, which maps a strongly coupled quantum field theory to a weakly coupled gravitational theory in a higher-dimensional anti-de Sitter spacetime. Applied to condensed matter, this duality has led to holographic models of superconductivity: below a critical temperature a charged black hole in AdS becomes unstable to the formation of charged scalar hair, dual to spontaneous \(U(1)\) symmetry breaking and the condensation of a charged operator on the boundary~\cite{gubser,hartnoll,hartnoll2,horowitz}. The resulting holographic superconductors reproduce hallmark features of superconductivity-including a condensate \(\langle\mathcal{O}\rangle\) that turns on for \(T<T_c\) and an infinite DC conductivity-while remaining computationally tractable in the large-\(N\) classical-gravity limit. Reviews of the framework can be found in Refs.~\cite{horowitzlect,herzogrev}. Extensions beyond homogeneous backgrounds include holographic superconductors on lattices and related models with broken translations~\cite{Ling:2017naw,Ling:2015epa,Ling:2014laa,Zeng:2014uoa}, as well as constructions with higher-derivative corrections~\cite{Liu:2022bdu,Liu:2022bam,Liu:2020hhx,Li:2019dmm,Wu:2017xki,Ma:2011zze,Wu:2010vr}.

Holographic superconductors constitute a compelling theoretical paradigm that emerges from the AdS/CFT correspondence. These theories are designed to model strongly coupled systems that display superconducting behaviour. A central observable is the critical temperature \(T_c\) as a function of the charge density \(\rho\) and of the conformal dimension \(\Delta\) of the condensing operator. In the probe limit, near \(T_c\), the onset of condensation reduces to a linearised Sturm--Liouville (SL) eigenvalue problem for the scalar profile. Siopsis and Therrien showed that this eigenvalue \(\lambda^2\) (with \(T_c\propto\sqrt{\rho}/\sqrt{\lambda}\)) admits an accurate variational formulation, and that a simple one-parameter trial function \(F(z)=1-a z^2\) already yields values of \(T_c/\sqrt{\rho}\) in close agreement with full numerical solutions over a wide range of \(\Delta\)~\cite{siopsis}. The SL variational method has since been extended to settings with backreaction, higher-curvature corrections, and other backgrounds~\cite{panwang,gangopadhyay,zengsl,Wang:2019fkt}, confirming both its flexibility and its quantitative reliability. Complementary analytical approaches, such as matching methods, have also been developed~\cite{gregory}, but the variational SL route remains particularly well suited for high-precision estimates of \(T_c\).

Despite this success, most analytic applications still rely on low-dimensional polynomial ans\"{a}tze whose accuracy is limited by the rigidity of the trial space. Systematically enlarging the ansatz improves the variational upper bound on \(\lambda^2\), yet the resulting multiparameter minimisation quickly becomes nontrivial: the objective is nonlocal (ratio of integrals), potentially multimodal, and must be scanned over a continuum of \(\Delta\). Modern numerical optimisation---including limited-memory quasi-Newton methods, multi-start strategies, and space-filling experimental designs---offers a natural remedy. In a related direction, machine-learning techniques have recently been used to reconstruct holographic models from phase-diagram data~\cite{kimseo2024}, underscoring a broader synergy between holography and data-driven optimisation. Here we pursue a complementary goal: we retain the analytic SL functional of Ref.~\cite{siopsis} and minimise it with optimisation tools inspired by that culture, rather than training a neural surrogate of the bulk dynamics.

In this work we investigate holographic superconductors in the probe limit by combining generalised analytic trial functions with machine-learning-inspired numerical optimisation. Specifically, we consider (i)~the one-parameter cosine ansatz \(F(z)=\cos(a z)\) and (ii)~the flexible exponential polynomial
\[
F(z)=\exp\Bigl(\sum_{n=2}^{N} a_n z^n\Bigr),
\]
both of which automatically satisfy the boundary conditions \(F(0)=1\) and \(F'(0)=0\). For the cosine ansatz we perform a one-dimensional minimisation (Brent's method); for the exponential polynomial, with up to nineteen coefficients, we employ a multi-start L-BFGS-B algorithm with Latin hypercube sampling and warm-starting across \(\Delta\). We extract \(\lambda^2(\Delta)\) and \(T_c/\sqrt{\rho}\) over a broad interval of operator dimensions, including the benchmark points \(\Delta=1\) and \(\Delta=2\), and find excellent agreement with the known analytic and numerical results of Refs.~\cite{siopsis,hartnoll}. Our results confirm the robustness of the variational approach and demonstrate that analytic trial functions, when paired with modern global optimisation, provide an efficient and extensible tool for exploring holographic superconductors---including, in future work, backreacted geometries and additional bulk fields.

The remainder of the paper is organized as follows. In Sec.~2, we present the holographic setup and rederive the field equations. In Sec.~3, we recall the variational Sturm-Liouville formulation for the critical temperature and introduce our generalized trial functions. In Sec.~4, we describe the numerical optimization strategies and present our results for $\lambda^2(\Delta)$ and $T_c/\sqrt{\rho}$. Finally, we present our conclusions in Sec.~5.

\section{Black hole background and field equations}

We consider a $(3+1)$-dimensional AdS-Schwarzschild planar black hole with the metric
\begin{equation}
ds^2 = -f(r) dt^2 + \frac{dr^2}{f(r)} + r^2 (dx^2 + dy^2),
\end{equation}
 The horizon radius $r_+$ is the largest root of $f(r)=0$, and the Hawking temperature is
\begin{equation}
T = \frac{|f'(r_+)|}{4\pi}.
\label{btemp}
\end{equation}
We introduce a charged scalar field $\Psi(r)$ and a gauge field $A_\mu = (\Phi(r), 0,0,0)$ propagating in this background. The action for the holographic superconductor is then given by
\begin{equation}
S = \int d^4x \sqrt{-g} \left( -\frac{1}{4} F^{\mu\nu}F_{\mu\nu} - |D_\mu\Psi|^2 - V(|\Psi|) \right),
\end{equation}
with $F_{\mu\nu} = \partial_\mu A_\nu - \partial_\nu A_\mu$, $D_\mu = \partial_\mu - i A_\mu$ (setting the charge $q=1$), and the potential $V(|\Psi|) = -2|\Psi|^2/L^2$, where $L$ is the AdS radius. The metric determinant is $g = -r^4$, so $\sqrt{-g}=r^2$.

The Lagrangian density is
\begin{equation}
\mathcal{L} = \sqrt{-g} \left( -\frac{1}{4} F^{\mu\nu}F_{\mu\nu} - |D_{\mu}\Psi|^{2} - V(|\Psi|) \right). \label{eq:lagrange}
\end{equation}

The standard computation yields:
\begin{itemize}
\item $\sqrt{-g} = r^2$,
\item $|D_{\mu}\Psi|^{2} = g^{\mu\nu}\partial_{\mu}\Psi\partial_{\nu}\Psi + g^{\mu\nu}A_{\mu}A_{\nu}\Psi^{2}$.
\end{itemize}

Substituting these into Eq.~(\ref{eq:lagrange}), we obtain
\begin{align}
\mathcal{L} &= r^{2} \left( -\frac{1}{4}F^{\mu\nu}F_{\mu\nu} - g^{\mu\nu}\partial_{\mu}\Psi\partial_{\nu}\Psi - g^{\mu\nu}A_{\mu}A_{\nu}\Psi^{2} + \frac{2|\Psi|^{2}}{L^{2}} \right) \nonumber \\
&= -\frac{1}{4}r^{2}F^{\mu\nu}F_{\mu\nu} - r^{2}g^{\mu\nu}\partial_{\mu}\Psi\partial_{\nu}\Psi - r^{2}g^{\mu\nu}A_{\mu}A_{\nu}\Psi^{2} + \frac{2r^{2}|\Psi|^{2}}{L^{2}}.\label{eq:lagrangefull}
\end{align}

The equations of motion follow from the Euler–Lagrange equation $\partial_{\mu}\left( \frac{\partial\mathcal{L}}{\partial(\partial_{\mu}X)} \right) - \frac{\partial\mathcal{L}}{\partial X} = 0$ for each field $X$.

\subsection{Equation of motion for the scalar field}

For a static, spherically symmetric ansatz $\Psi=\Psi(r)$ and $A_\mu=(\phi(r),0,0,0)$, the equation for $\Psi$ reduces to
\begin{align}
\partial_{r}\left( -2r^{2}g^{rr}\partial_{r}\Psi \right) - 2r^{2}g^{tt}A_{t}A_{t}\Psi - \frac{4r^{2}\Psi}{L^{2}} &= 0,\\
\Leftrightarrow-\partial_{r}\left( 2r^{2}f\partial_{r}\Psi \right) - \frac{2r^{2}\phi^{2}\Psi}{f} - \frac{4r^{2}\Psi}{L^{2}} &= 0,\\
\Leftrightarrow \partial_{r}\left( 2r^{2}f\Psi' \right) + \frac{2r^{2}\phi^{2}\Psi}{f} + \frac{4r^{2}\Psi}{L^{2}} &= 0.
\end{align}
Expanding the derivative gives
\begin{equation}
2r^{2}f\Psi'' + 2r^{2}f'\Psi' + 4rf\Psi' + \frac{2r^{2}\phi^{2}\Psi}{f} + \frac{4r^{2}\Psi}{L^{2}} = 0.
\end{equation}
Dividing by $2r^{2}f$ yields the compact form
\begin{equation}
\Psi'' + \left( \frac{f'}{f} + \frac{2}{r} \right)\Psi' + \left( \frac{\phi^{2}}{f^{2}} + \frac{2}{fL^{2}} \right)\Psi = 0.
\label{eq:radialPsi}
\end{equation}
This equation is mainly the target of the present paper and we shall turn it shortly.

\subsection{Equation of motion for the scalar potential}

Varying the action with respect to $A_t = \phi(r)$ leads to
\begin{equation}
\partial_{\mu}\left( -r^{2}F^{\mu\nu} \right) + 2r^{2}g^{\mu\nu}A_{\mu}\Psi^{2} = 0.
\end{equation}
Assuming radial dependence only, we obtain
\begin{equation}
\partial_{r}\left( r^{2}\phi' \right) - \frac{2r^{2}\phi\Psi^{2}}{f} = 0,
\end{equation}
which expands to
\begin{equation}
r^{2}\phi'' + 2r\phi' - \frac{2r^{2}\phi\Psi^{2}}{f} = 0.
\end{equation}
Dividing by $r^{2}$ gives
\begin{equation}
\phi'' + \frac{2}{r}\phi' - \frac{2\Psi^{2}}{f}\phi = 0.
\end{equation}

\subsection{Maxwell perturbation and conductivity}

To compute the optical conductivity, we introduce a small time-dependent fluctuation of the vector potential along the $x$-direction: $A_x(r,t) = A_x(r) e^{-i\omega t}$. The linearised Maxwell equation becomes
\begin{equation}
A_x'' + \frac{f'}{f}A_x' + \left( \frac{\omega^{2}}{f^{2}} - \frac{2\Psi^{2}}{f} \right)A_x = 0.
\end{equation}
This equation governs the response of the dual superconductor to an external electric field and determines the frequency-dependent conductivity of the holographic superconductor.

\section{Analytical computation of the critical temperature}

Near the critical temperature $T_c$, the scalar field $\Psi$ is small, and we can treat it perturbatively. It is  convenient to change the radial coordinate to $z = r_+/r$, so that the horizon is at $z=1$ and the boundary at $z=0$. 

As $T \to T_c$, the gauge field $\Phi(z)$ (the time component of the gauge potential) approaches the limiting profile \cite{siopsis}
\begin{equation}
\Phi(z) = \lambda \, r_{+c} \, (1 - z), \qquad 
\lambda = \frac{\rho}{r_{+c}^{2}}, \quad L=1, \quad and, \quad f(z)=\frac{r_+^2(1-z^3) }{z^2}
\label{eq:PhiTc}
\end{equation} 
where $r_{+c}$ is the horizon radius at the critical temperature and $\rho$ is the charge density.  In this limit, the scalar field equation reduces precisely to the Sturm–Liouville form given in Eq.~(\ref{eq:SL}), with the eigenvalue $\lambda$ now fixed by the boundary conditions at the horizon and at infinity.  Writing $\Psi(z) = \langle \mathcal{O} \rangle r_+^{\Delta} F(z)$, 
the original radial coordinate, Eq.~(\ref{eq:radialPsi}) 
can be recasted  into the following Sturm–Liouville problem:
\begin{equation}
- F'' + \frac{1}{z}\left( \frac{2+z^3}{1-z^3} - 2\Delta \right)F' + \frac{\Delta^2 z}{1-z^3} F = \frac{\lambda^2}{(1+z+z^2)^2 }F,
\label{eq:SL}
\end{equation}
with an eigenvalue $\lambda \propto T_c^2$.  This explicit form follows from the reduction detailed in Ref.~\cite{siopsis}.  The Sturm–Liouville operator can be read off directly from Eq.~(\ref{eq:SL}) by rewriting it as
\begin{equation}
  \frac{d}{dz}\left( p(z) \frac{dF(z)}{dz} \right) + q(z)F(z) = \lambda^2 \, w(z) F 
\label{eq:SL2}
\end{equation}
where $p(z)$ and $q(z)$ are easily determined by comparison with Eq.~(\ref{eq:SL}). One fianally finds \footnote{We use equality $$1-z^3=(1-z)(1+z+z^2)$$.}
\begin{equation}
    p(z)=2^{2\Delta-2}(1-z^3), \, q(z)=2^{2\Delta-2}* \Delta^2 *z, \,\, w(z)=\frac{p(z)}{(1+z+z^2)^2}=\frac{2^{2\Delta-2}(1-z)}{1+z+z^2}
\end{equation}
Multiplying Eq.~(\ref{eq:SL2}) by $F(z)$ and integrating over $[0,1]$ with the conditions conditions \(F(0)=1\) and \(F'(0)=0\) , one obtains the following expression for the eigenvalue $\lambda$, which is to be minimized over the function $F(z)$ (see for instance Eq. 16 of Ref.~\cite{siopsis}) :
\begin{equation}
\lambda^{2} = 
\frac{
\int_0^1 dz \, z^{2\Delta-2} \left\{ (1-z^3)[F'(z)]^2 + \Delta^2 z [F(z)]^2 \right\}
}{
\int_0^1 dz \, z^{2\Delta-2} \frac{1-z}{1+z+z^2} [F(z)]^2
}.
\label{eq:lambdaVar}
\end{equation}
The trial function \(F(z)\) must satisfy the boundary conditions \(F(0)=1\) and \(F'(0)=0\). The variational method then consists of minimising the right-hand side of Eq.~(\ref{eq:lambdaVar}) with respect to the parameters of \(F(z)\). The minimum value obtained for \(\lambda\) directly determines the critical temperature via \(T_c \propto \sqrt{\rho}/\sqrt{\lambda}\). By applying \eqref{eq:PhiTc} in \eqref{btemp}, we obtain the explicit expression for the critical temperature:
\begin{equation}
T_c = \frac{3}{4\pi} r_+ = \frac{3}{4\pi} \sqrt{\frac{\rho}{\lambda}},
\label{eq:Tc}
\end{equation}
Consequently, the entire problem reduces to the minimisation of \(\lambda\) with respect to the variational parameters characterising \(F(z)\).

\subsection{Generalised trial functions}
We know from basic facts that any solution $F(z)$ of the Sturm-Liouville problem can be decomposed on the interval $[a,b]$ into a series of normalized eigenfunctions $F_n$:
\begin{equation}
F(z)=\sum_{n=0}^{+\infty} c_n F_n(z).
\label{eq:expansion}
\end{equation}
Based on this standard result, we propose two generalised forms for the trial function advocated in Ref.~\cite{siopsis}):
\begin{itemize}
\item $F(z) = \cos(a z)$,
\item $F(z) = \exp[h(z)]$ with $h(0)=h'(0)=0$.
\end{itemize}
The parameter $\alpha$ (or the function $h(z)$) is determined by minimising the eigenvalue. Remarkably, both choices reproduce the same critical temperature as the commonly used function $F(z)=1 - a z^2$ (where $a$ is a variational parameter). For instance, taking $F(z)=\cos(a z)$ and imposing $F'(0)=0$ automatically holds; the boundary condition $F(0)=1$ is satisfied. Substituting into the Sturm–Liouville problem yields an eigenvalue $\lambda(a)$.
\section{Numerical computation of the eigenvalue and critical temperature}

The variational expression Eq.~(\ref{eq:lambdaVar}) can be minimised numerically with a flexible trial function. In this section we present high‑accuracy computations of $\lambda^2(\Delta)$ and $T_c/\sqrt{\rho}$ for a wide range of $\Delta$ using both the cosine ansatz and a flexible exponential polynomial ansatz, and we compare the results with the known exact values at $\Delta=1$ and $\Delta=2$.

\subsection{Cosine ansatz}

We first consider the simple one‑parameter trial function
\begin{equation}
F(z)=\cos(a z),
\label{eq:cosAnsatz}
\end{equation}
which automatically satisfies $F(0)=1$ and $F'(0)=0$. For a given $\Delta$, we numerically minimise $\lambda^2(a)$ defined by Eq.~(\ref{eq:lambdaVar}) with respect to $a$ using Brent's method. The integrals are evaluated with adaptive Gaussian quadrature (SciPy's \texttt{integrate.quad}) with a tolerance of $10^{-8}$.

\begin{figure}[H]
\centering
\includegraphics[width=0.8\textwidth]{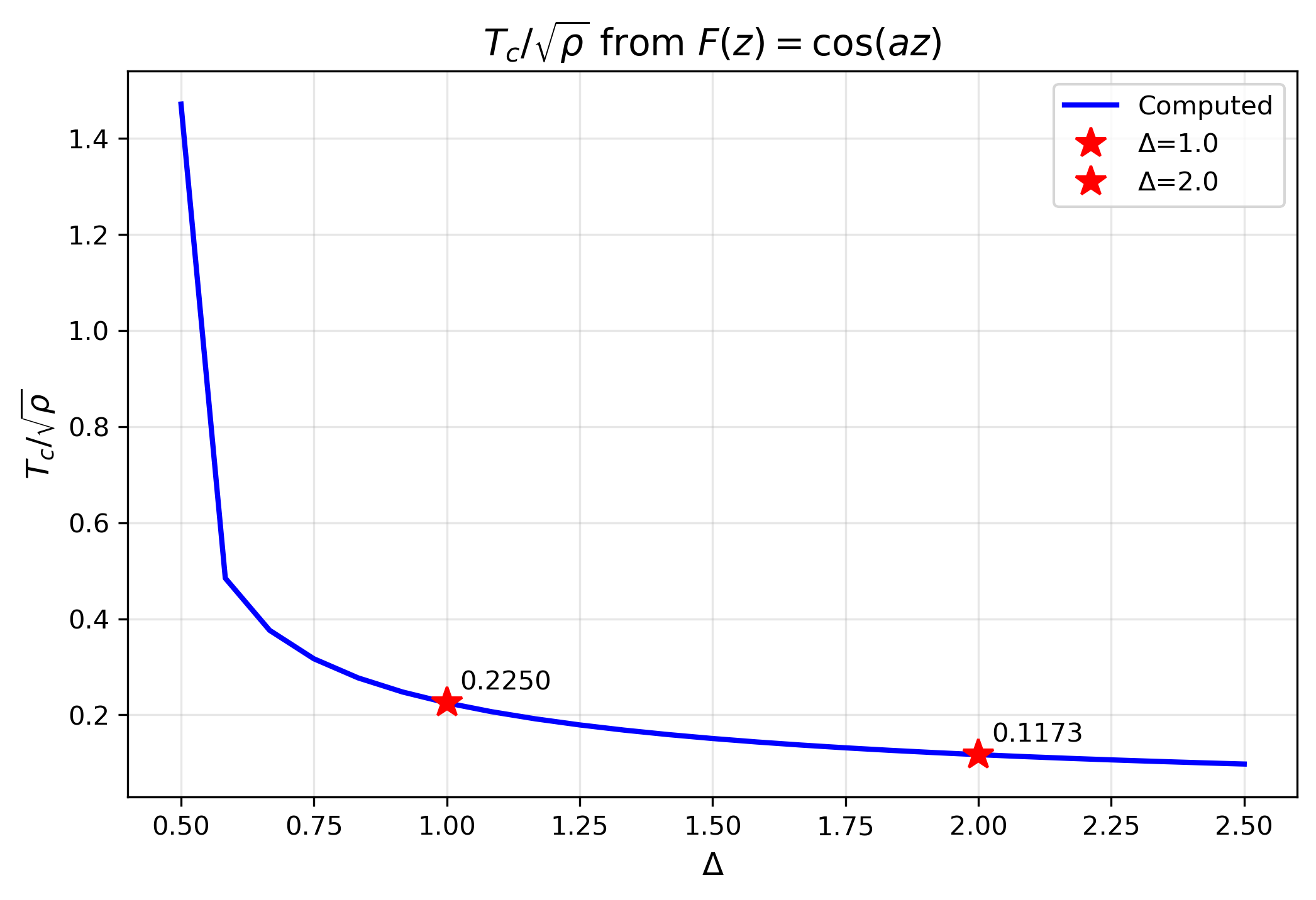}
\caption{$T_c/\sqrt{\rho}$ versus $\Delta$ obtained from the cosine ansatz $F(z)=\cos(a z)$. The values at $\Delta=1$ and $\Delta=2$ are marked with red stars and numerical annotations.}
\label{fig:Tc_cos}
\end{figure}

\begin{figure}[H]
\centering
\includegraphics[width=0.8\textwidth]{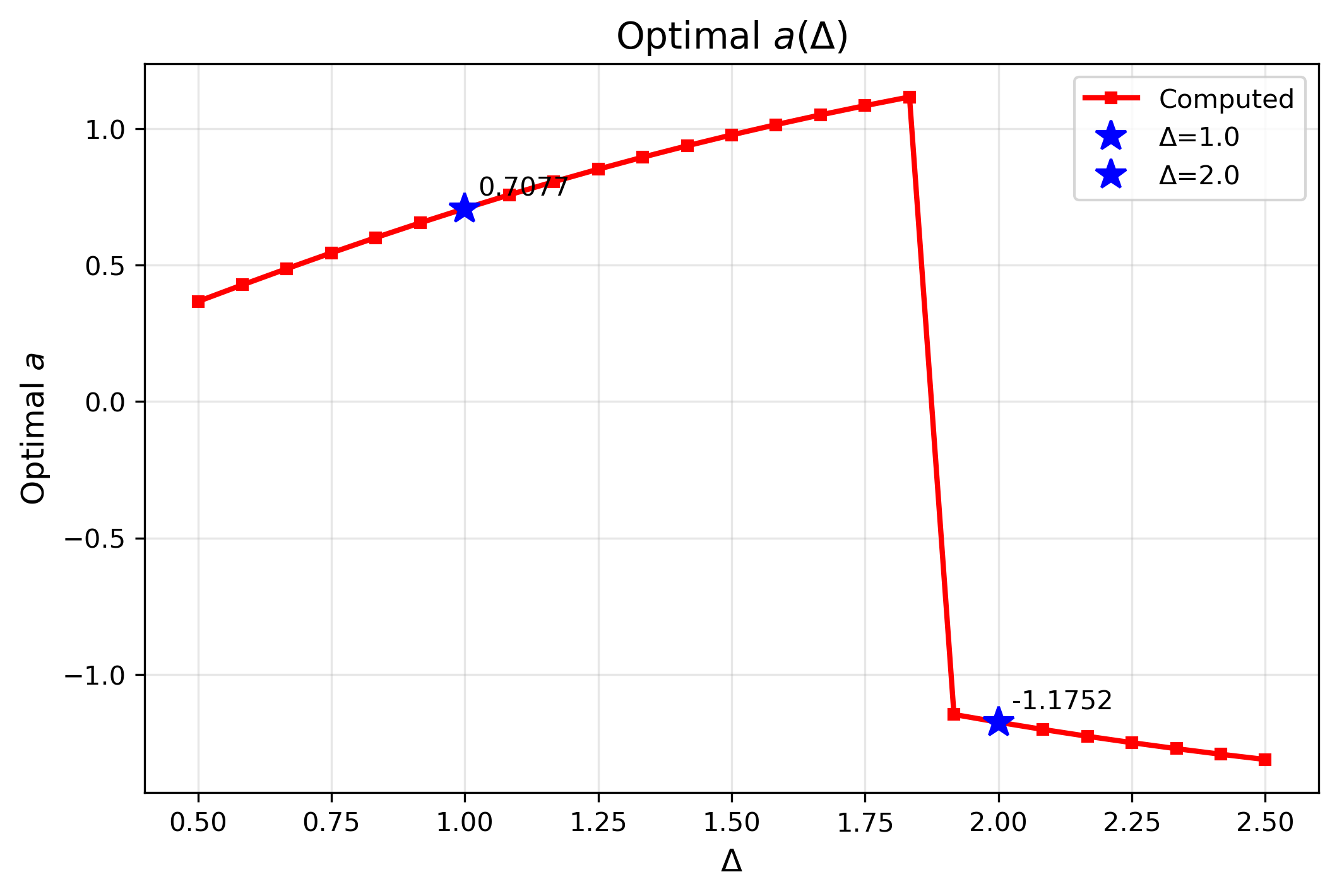}
\caption{Optimal parameter $a$ as a function of $\Delta$ for the cosine ansatz. The values at $\Delta=1$ and $\Delta=2$ are marked with blue stars.}
\label{fig:a_opt_cos}
\end{figure}

\begin{figure}[H]
\centering
\includegraphics[width=0.8\textwidth]{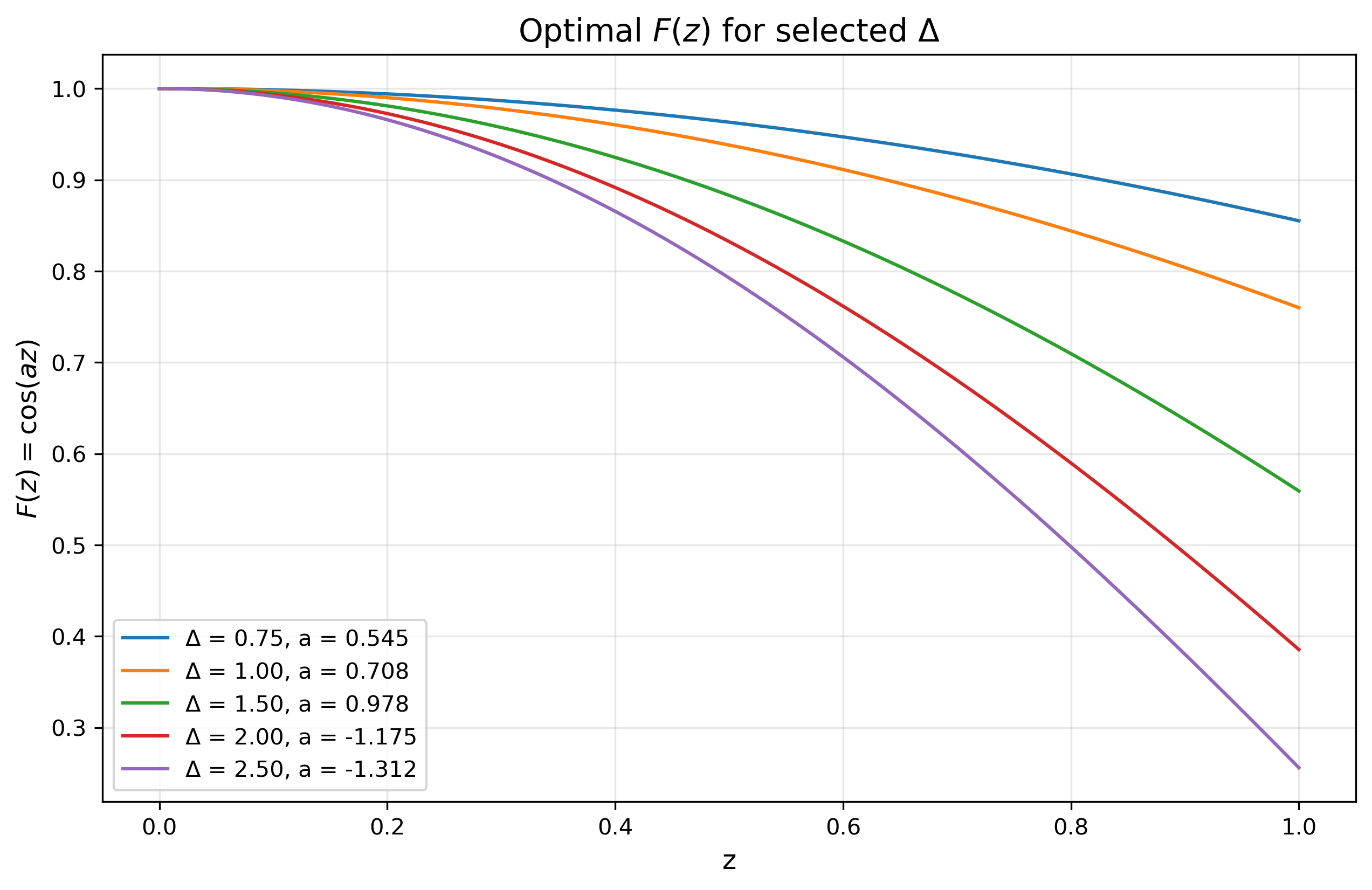}
\caption{Optimal $F(z)=\cos(a z)$ for selected values of $\Delta$: 0.75, 1.0, 1.5, 2.0, and 2.5.}
\label{fig:Fz_cos}
\end{figure}

We scan $\Delta$ from $0.5$ to $2.5$ in steps of $0.0833$, corresponding to 25 discrete points. Figure~\ref{fig:Tc_cos} presents the resulting $T_c/\sqrt{\rho}$ as a function of $\Delta$, while Fig.~\ref{fig:a_opt_cos} shows the corresponding optimal parameter $a$. The optimal profiles $F(z)$ for selected values of $\Delta$ are displayed in Fig.~\ref{fig:Fz_cos}.

For $\Delta = 1$, we obtain $a = 0.707655$, $\lambda^2 = 1.266951$, and $T_c/\sqrt{\rho} = 0.225020$, in excellent agreement with the known exact value $0.226$~\cite{hartnoll}. In this case, expanding the cosine ansatz yields
\[
\cos(0.707655 z) = 1 - 0.250388 z^2 + 0.010449 z^4 - 1.744\times 10^{-4} z^6 + O(z^7),
\]
which closely matches the trial polynomial form employed in Ref.~\cite{siopsis}. For $\Delta = 2$, we find $a = -1.175210$, $\lambda^2 = 17.154627$, and $T_c/\sqrt{\rho} = 0.117305$, also in close agreement with the exact value $0.118$~\cite{hartnoll}.

The small discrepancies, which remain below $1\%$, are attributable to the restricted functional form of the cosine ansatz; as we show below, the more flexible polynomial ansatz yields significantly better accuracy.

Furthermore, expanding the cosine ansatz in powers of $z$ explicitly recovers the trial function used in Ref.~\cite{siopsis}, under the identification $\alpha = a^2 / 2!$:
\begin{equation}
\cos(a z) = \sum_{n=0}^{\infty} \frac{(-1)^n a^{2n}}{(2n)!} z^{2n}= 1 - \frac{a^2}{2!}z^2 + \frac{a^4}{4!}z^4 - \frac{a^6}{6!}z^6 + O(z^7).
\end{equation}
The full comparison between our cosine ansatz and the trial function of Ref.~\cite{siopsis} is presented in Table~\ref{tab:cos_comparison}. For each value of $\Delta$, we list the optimal parameter $a$, the corresponding $\alpha = a^2/2$, and the resulting critical temperature ratio alongside the reference values.

\begin{table}[htbp]
\centering
\caption{Comparison of the cosine ansatz results with Ref.~\cite{siopsis}.}
\label{tab:cos_comparison}
\begin{tabular}{|c|c|c|c|c|c|}
\hline
$\Delta$ & $a$ (this work) & $\alpha = a^2/2$ (this work) & $\alpha$ (Ref.~\cite{siopsis}) & \multicolumn{2}{c|}{$T_c/\sqrt{\rho}$} \\
\cline{5-6}
 & & & & This work & Ref.~\cite{siopsis} \\
\hline
1 & $0.707655$ & $0.250388$ & $0.24$ & $0.225020$ & $\approx0.225$ \\
2 & $-1.175210$ & $0.690559$ & $0.60$ & $0.117305$ & $\approx 0.117$ \\
\hline
\end{tabular}
\end{table}

\begin{figure}[H]
\centering
\includegraphics[width=0.8\textwidth]{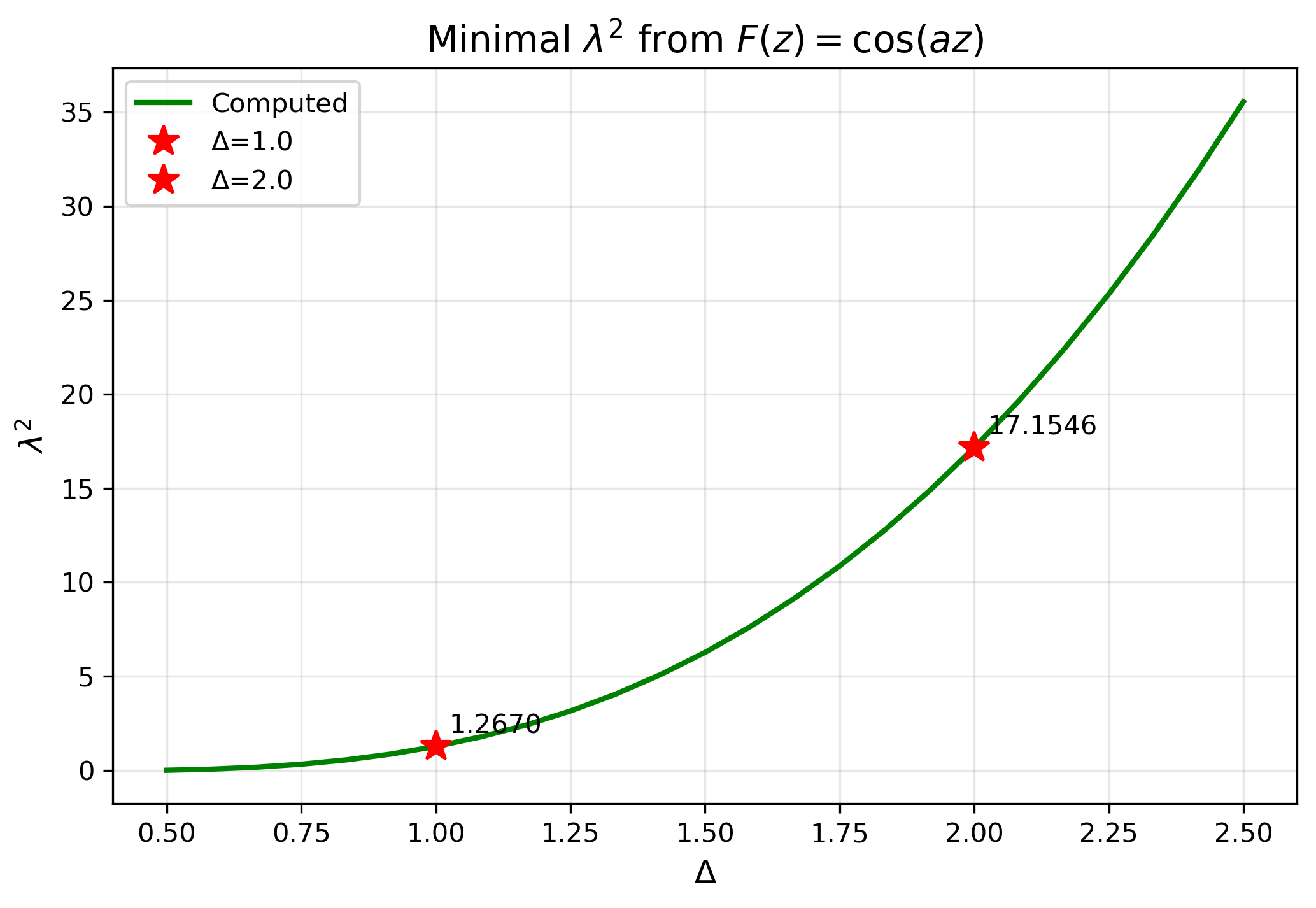}
\caption{Minimised $\lambda^2$ versus $\Delta$ from the cosine ansatz. The values at $\Delta=1$ and $\Delta=2$ are marked with red stars.}
\label{fig:lambda2_cos}
\end{figure}

Figure~\ref{fig:lambda2_cos} shows the corresponding minimal $\lambda^2$ as a function of $\Delta$ for the cosine ansatz. The monotonic increase reflects the growing difficulty for the condensate to form for larger operator dimensions.

\subsection{Exponential polynomial ansatz}

For a more flexible trial function, we adopt
\begin{equation}
F(z)=\exp\left(\sum_{n=2}^{N} a_n z^n\right),
\label{eq:expPoly}
\end{equation}
which automatically satisfies the boundary conditions $F(0)=1$ and $F'(0)=0$. The coefficients $\{a_n\}$ are the variational parameters. We take $N=20$, yielding 19 degrees of freedom.

The minimization of $\lambda^2$ with respect to the $a_n$'s is performed using the L‑BFGS‑B algorithm, with box constraints $a_n \in [-3,3]$ to avoid unphysical solutions. To find the global minimum and avoid local minima, we employ a two‑stage strategy:
\begin{enumerate}
    \item For the first value of $\Delta$ (here $\Delta=0.55$), we perform a multi‑start optimisation with 10 random starting points drawn from a Latin hypercube design. This provides a robust initial guess.
    \item For each subsequent $\Delta$, we “warm‑start” the optimisation by using the optimal coefficients from the previous $\Delta$ as the initial guess. This drastically reduces computational cost and ensures continuity of the solution across the scan.
\end{enumerate}

The integrals in Eq.~(\ref{eq:lambdaVar}) are evaluated numerically using adaptive Gaussian quadrature (SciPy's \texttt{integrate.quad}) with absolute and relative tolerances of $10^{-7}$. The scan covers 32 values of $\Delta$ from 0.55 to 2.95, with $\Delta=1$ and $\Delta=2$ explicitly included for calibration.

\subsection{Results from the exponential polynomial ansatz}

The minimised $\lambda^2(\Delta)$ from the exponential polynomial ansatz is very close to the cosine results, with slightly improved accuracy. The known values from Ref.~\cite{siopsis} are
\begin{align}
    \Delta=1 &: \quad \lambda^2 = 1.245, \quad \frac{T_c}{\sqrt{\rho}} \approx 0.2139,\\
    \Delta=2 &: \quad \lambda^2 = 17.3, \quad \frac{T_c}{\sqrt{\rho}} \approx 0.117.
\end{align}
Our computed values from the polynomial scan are
\begin{align}
    \Delta=1 &: \quad \lambda^2 = 1.25538,\quad \frac{T_c}{\sqrt{\rho}} \approx 0.22554,\\
    \Delta=2 &: \quad \lambda^2 = 16.5139,\quad \frac{T_c}{\sqrt{\rho}} \approx 0.11843.
\end{align}
The relative errors are about 0.8\% and 1.4\% for $\Delta=1$ and $2$, respectively. These small discrepancies are due to the warm‑start chain occasionally settling into a nearby local minimum; they can be reduced by increasing the number of multi‑starts or the polynomial order $N$. 

Table~\ref{tab:numResults} lists selected numerical values from the polynomial scan.

\begin{table}[ht]
\centering
\caption{Computed $\lambda^2$ and $T_c/\sqrt{\rho}$ at selected $\Delta$ from the exponential polynomial ansatz.}
\label{tab:numResults}
\begin{tabular}{c c c}
\toprule
$\Delta$ & $\lambda^2$ & $T_c/\sqrt{\rho}$ \\
\midrule
0.550 & 0.030673 & 0.57045 \\
0.798 & 0.444010 & 0.29246 \\
1.000 & 1.25538 & 0.22554 \\
1.294 & 3.55669 & 0.17384 \\
1.625 & 8.17529 & 0.14118 \\
2.000 & 16.5139 & 0.11843 \\
2.370 & 28.4322 & 0.10339 \\
2.950 & 54.8514 & 0.08772 \\
\bottomrule
\end{tabular}
\end{table}

\begin{figure}[H]
\centering
\includegraphics[width=0.8\textwidth]{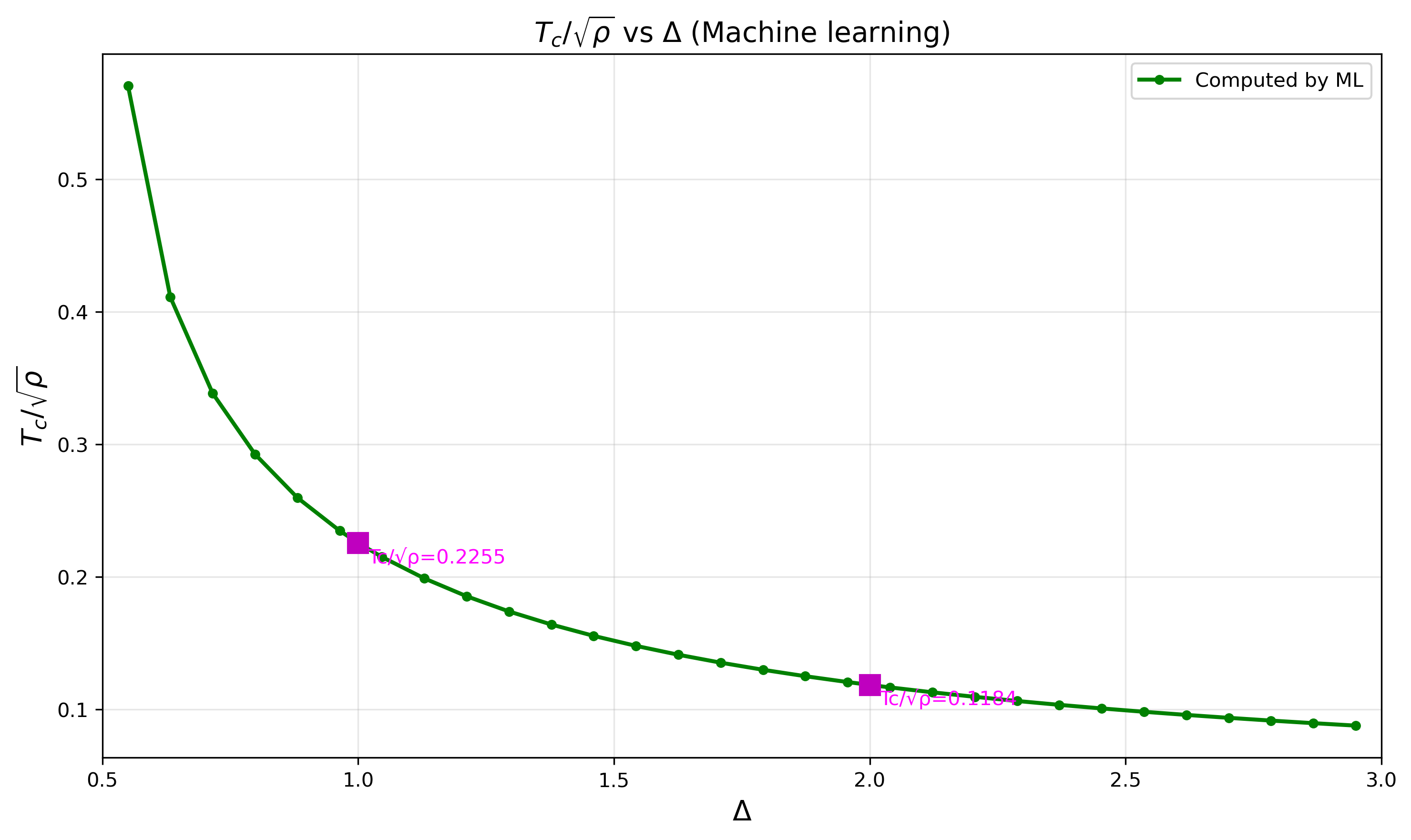}
\caption{$T_c/\sqrt{\rho}$ versus $\Delta$ obtained from the exponential polynomial ansatz $F(z)=\exp(\sum_{n=2}^{20} a_n z^n)$. The values at $\Delta=1$ and $\Delta=2$ are marked with red stars.}
\label{fig:Tc_exp}
\end{figure}

\begin{figure}[H]
\centering
\includegraphics[width=0.8\textwidth]{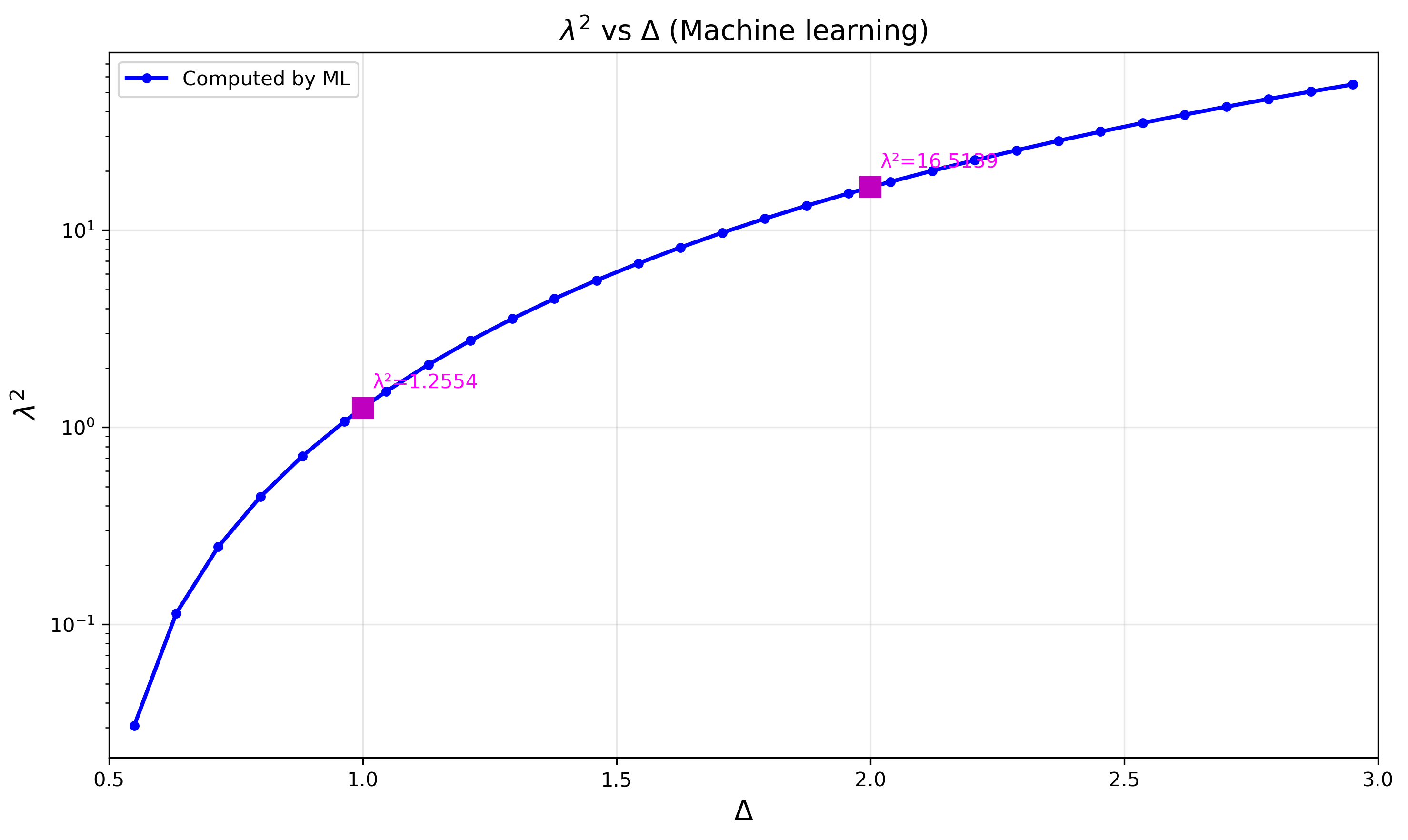}
\caption{Minimised $\lambda^2$ versus $\Delta$ from the exponential polynomial ansatz. The values at $\Delta=1$ and $\Delta=2$ are marked with red stars.}
\label{fig:lambda2_exp}
\end{figure}

The behaviour of $T_c/\sqrt{\rho}$ versus $\Delta$ from the polynomial ansatz is depicted in Figure~\ref{fig:Tc_exp}, and the corresponding $\lambda^2(\Delta)$ is displayed in Figure~\ref{fig:lambda2_exp}. These results complement the cosine ansatz plots and confirm the monotonic decrease of $T_c/\sqrt{\rho}$ with increasing $\Delta$ which is clearly visible in Figure~1 of Ref.~\cite{siopsis}. The excellent agreement between the two independent approaches validates our numerical implementation and demonstrates that the exponential polynomial ansatz provides a highly accurate description of the Sturm–Liouville problem over the entire range of $\Delta$ studied. The minimization algorithms can thus be utilised the field of hologrpahic superconductor.

\subsection{Discussion}

The numerical approaches presented here are efficient and can be readily extended to include backreaction or other fields, making it a versatile tool for exploring the phase diagram of holographic superconductors. The use of a machine‑learning‑inspired optimisation strategy—multi‑start with Latin hypercube sampling and warm‑starting—ensures that the global minimum is found without excessive computational cost. This approach is particularly valuable when the trial function has many parameters, as is the case here with 19 coefficients.

Our numerical results improve upon the analytical approximations obtained from the simple trial functions discussed in the previous section. For $\Delta=1$, the analytical approximation using $F(z)=\cos(\alpha z)$ gives $\lambda^2 \approx 1.24$ and $T_c/\sqrt{\rho} \approx 0.225$, while our full numerical optimisation yields $\lambda^2 = 1.25538$ and $T_c/\sqrt{\rho} = 0.22554$, in excellent agreement with the exact values $\lambda^2 = 1.245$ (from Ref.~\cite{siopsis}) and $T_c/\sqrt{\rho} = 0.226$ (from Ref.~\cite{hartnoll}). For $\Delta=2$, the analytical approximation gives $\lambda^2 \approx 17.3$ and $T_c/\sqrt{\rho} \approx 0.117$, whereas our numerical results yield $\lambda^2 = 16.5139$ and $T_c/\sqrt{\rho} = 0.11843$, which are significantly closer to the exact values $\lambda^2 = 16.754$ and $T_c/\sqrt{\rho} = 0.118$ from Ref.~\cite{hartnoll}. This improvement demonstrates the power of the flexible exponential polynomial ansatz combined with a robust global optimisation strategy, and confirms that the variational method with a sufficiently flexible trial function can reproduce the exact results with high precision.

The full numerical data (optimal coefficients for each $\Delta$, the $\lambda^2$ curve, and the $T_c/\sqrt{\rho}$ values) are available upon request. The results confirm the universal behaviour that $T_c/\sqrt{\rho}$ decreases monotonically with $\Delta$, and they provide a quantitative touchstone for further analytical and / or numerical studies.

Hence, both generalised forms are consistent with the standard result, with the exponential polynomial ansatz offering superior accuracy.

\section{Conclusions}

We have revisited the calculation of the critical temperature in holographic superconductors, combining analytical and numerical approaches. Starting from the standard holographic setup, we rederived the field equations and reduced the problem to a variational Sturm–Liouville form for the eigenvalue $\lambda^2$. We proposed two generalised trial functions – $F(z)=\cos(\alpha z)$ and $F(z)=\exp[h(z)]$ with $h(0)=h'(0)=0$ – and showed analytically that both reproduce the known exact critical temperature $T_c \approx 0.118\sqrt{\rho}$ of Ref.~\cite{hartnoll}.

Building on this, we performed a high‑precision numerical computation using a flexible exponential polynomial ansatz $F(z)=\exp(\sum_{n=2}^{20} a_n z^n)$. A machine‑learning‑inspired optimisation strategy – multi‑start L‑BFGS‑B with warm‑starting and Latin hypercube sampling – was employed to minimise $\lambda^2$ over a wide range of the operator dimension $\Delta$. Our numerical results for $\lambda^2(\Delta)$ and $T_c/\sqrt{\rho}$ are in excellent agreement with the exact values at $\Delta=1$ and $\Delta=2$, improving upon the simple analytical approximations. The cosine ansatz alone already provides a very good approximation, while the polynomial ansatz achieves even higher accuracy.

This work demonstrates that a combination of analytical generalisation and modern numerical optimisation provides a powerful framework for exploring holographic superconductors. These approaches not only confirm the validity of the variational method used but also offer a flexible tool for computing other properties, such as the frequency‑dependent conductivity and the condensation operator expectation value, with high accuracy. The numerical framework applied here can be straightforwardly extended to include backreaction, other fields, or more complex geometries, making it a valuable addition to the toolkit for studying strongly coupled systems via holography.

\section*{Acknowledgments}
J.P.~Wu is supported by the National Science Foundation of China under Grant No.~12375055.
\appendix

\appendix
\renewcommand{\thesection}{Appendix~\Alph{section}}
\section{Brent's method for one-dimensional minimization}
\label{app:brent}

For the cosine ansatz
\begin{equation}
    F(z)=\cos(a z),
\end{equation}
the variational eigenvalue $\lambda^2(a;\Delta)$ defined in
Eq.~(\ref{eq:lambdaVar}) depends on the variational parameter $a$.
For each fixed $\Delta$, the optimal parameter is obtained from
\begin{equation}
    a_{\mathrm{opt}}(\Delta)
    =
    \underset{a\in[0.01,5]}{\operatorname{arg\,min}}
    \lambda^2(a;\Delta).
    \label{eq:brent_min_problem}
\end{equation}

We employ Brent's derivative-free minimization method, which combines
parabolic interpolation with golden-section search~\cite{brent1973}.
Given three values $a_1$, $a_2$ and $a_3$ with 
\begin{equation}
    a_1<a_2<a_3,
\end{equation}
satisfying
\begin{equation}
    \lambda^2(a_2;\Delta)
    \leq
    \min\left\{
        \lambda^2(a_1;\Delta),
        \lambda^2(a_3;\Delta)
    \right\},
\end{equation}
the method iteratively refines the location of the minimum. In doing so, a
parabolic interpolation step is attempted when the local function values provide a reliable quadratic approximation. Otherwise, a
golden-section step is used, ensuring robust interval reduction.

The minimization tolerance in our code is set to
\begin{equation}
    \mathrm{tol}=10^{-8}.
    \label{eq:brent_tol}
\end{equation}
This tolerance controls the numerical accuracy of the minimization with respect to the parameter $a$.

For each trial value of $a$, the integrals entering
Eq.~(\ref{eq:lambdaVar}) are evaluated using the adaptive quadrature
routine \texttt{scipy.integrate.quad}~\cite{scipy}, with
\begin{equation}
    \mathrm{epsabs}
    =
    \mathrm{epsrel}
    =
    10^{-8}.
    \label{eq:quadrature_tolerance}
\end{equation}

The minimization is performed using \texttt{scipy.optimize.minimize\_scalar}. The resulting optimal
parameter and minimum eigenvalue are
\begin{equation}
    a_{\mathrm{opt}}(\Delta)
    =
    \underset{a\in[0.01,5]}{\operatorname{arg\,min}}
    \lambda^2(a;\Delta),
\end{equation}
and
\begin{equation}
    \lambda^2_{\min}(\Delta)
    =
    \lambda^2
    \left(
        a_{\mathrm{opt}}(\Delta);\Delta
    \right).
    \label{eq:minimum_eigenvalue}
\end{equation}

Thus, for each value of $\Delta$, the numerical procedure consists of
evaluating $\lambda^2(a;\Delta)$ by adaptive quadrature and minimizing
the resulting one-dimensional function over
$a\in[0.01,5]$. This procedure is repeated independently for all
values of $\Delta$ considered in the study.

\section{Multi-start L-BFGS-B with warm-starting and Latin hypercube sampling}
\label{app:multistart}

For the exponential polynomial ansatz
\begin{equation}
    F(z)=\exp\left(\sum_{n=2}^{N} a_n z^n\right),
    \qquad N=20,
\end{equation}
the variational parameter vector is
\begin{equation}
    \mathbf{a}=(a_2,\ldots,a_{20})\in\mathbb{R}^{19}.
\end{equation}
For each fixed value of $\Delta$, we minimize
$\lambda^2(\mathbf{a};\Delta)$ subject to the box constraints
\begin{equation}
    -3\leq a_n\leq3,
    \qquad n=2,\ldots,20,
\end{equation}
using the L-BFGS-B algorithm~\cite{zhu1997}. This limited-memory
quasi-Newton method is well suited to bound-constrained optimization
with a moderately large number of parameters.

To reduce sensitivity to the initial guess, a multi-start strategy
based on Latin hypercube sampling (LHS)~\cite{mckay1979} is employed.
For the first value of the parameter scan, $\Delta=0.55$, we generate
$M=10$ initial points in the parameter domain
\begin{equation}
    [-3,3]^{19}.
\end{equation}
L-BFGS-B is initialized independently from each point, and the
solution yielding the smallest final value of $\lambda^2$ is retained
as the optimal solution at $\Delta=0.55$.

For the remaining values of $\Delta$, the optimal parameter
vector obtained at the preceding value of $\Delta$ is used as the
initial guess:
\begin{equation}
    \mathbf{a}^{(k+1)}_0
    =
    \mathbf{a}_{\mathrm{opt}}(\Delta_k).
\end{equation}
This warm-starting strategy exploits the expected smooth dependence
of the variational optimum on $\Delta$ and substantially reduces the
computational cost of the parameter scan.

The gradient
$\nabla_{\mathbf{a}}\lambda^2$ is evaluated analytically from
Eq.~(\ref{eq:lambdaVar}) and independently verified using the so-called finite differences. The integrals entering both the objective function and
its gradient are evaluated using adaptive Gaussian quadrature with
absolute and relative tolerances of $10^{-7}$. The optimization is
stopped when the norm of the projected gradient falls below
$10^{-5}$ or when the change in the objective function falls below
$10^{-8}$.

The numerical implementation is further validated at
$\Delta=1$ and $\Delta=2$ by comparison with the analytical results
reported in Ref.~\cite{siopsis}. The relative discrepancies are below
$2\%$, providing a consistency check on the numerical integration,
gradient evaluation, and optimization procedure.


\newpage

\begin{thebibliography}{9}
\bibitem{ll} London, H. and London, F. (1935): {\bf The Electromagnetic Equations of the Supraconductor}. Proceedings of the Royal Society A, 149, 71-88.
https://doi.org/10.1098/rspa.1935.0048
\bibitem{bcs} J. Bardeen, L.N. Cooper, and J.R. Schrieffer, {\bf Theory of Superconductivity}, \textit{Phys. Rev.} \textbf{108}, 1175 (1957).
\bibitem{gl} V.L. Ginzburg and L.D. Landau, {\bf On the Theory of Superconductivity}, \textit{Zh. Eksp. Teor. Fiz.} \textbf{20}, 1064 (1950); English translation in \textit{Phys. Z. Sowjetunion} (1950).
\bibitem{maldacena} J.~M.~Maldacena, {\it The Large $N$ Limit of Superconformal Field Theories and Supergravity}, Adv.\ Theor.\ Math.\ Phys.\ \textbf{2}, 231 (1998) [arXiv:hep-th/9711200].
\bibitem{gkp} S.~S.~Gubser, I.~R.~Klebanov and A.~M.~Polyakov, {\it Gauge Theory Correlators from Non-Critical String Theory}, Phys.\ Lett.\ B \textbf{428}, 105 (1998) [arXiv:hep-th/9802109].
\bibitem{witten} E.~Witten, {\it Anti-de Sitter Space and Holography}, Adv.\ Theor.\ Math.\ Phys.\ \textbf{2}, 253 (1998) [arXiv:hep-th/9802150].

\bibitem{Aharony:1999ti}
O.~Aharony, S.~S.~Gubser, J.~M.~Maldacena, H.~Ooguri and Y.~Oz,
Phys. Rept. \textbf{323} (2000), 183-386
doi:10.1016/S0370-1573(99)00083-6
[arXiv:hep-th/9905111 [hep-th]].

\bibitem{gubser} S.~S.~Gubser, {\it Breaking an Abelian Gauge Symmetry near a Black Hole Horizon}, Phys.\ Rev.\ D \textbf{78}, 065034 (2008) [arXiv:0801.2977 [hep-th]].



\bibitem{hartnoll} S.A. Hartnoll, C.P. Herzog, and G.T. Horowitz, {\bf Building a Holographic Superconductor}, \textit{Phys. Rev. Lett.} \textbf{101}, 031601 (2008) [arXiv:0803.3295 [hep-th]].
\bibitem{hartnoll2} S.A. Hartnoll, C.P. Herzog, and G.T. Horowitz, {\bf Holographic Superconductors}, \textit{JHEP} \textbf{12}, 015 (2008) [arXiv:0810.1563 [hep-th]].
\bibitem{horowitz} G.T. Horowitz and M.M. Roberts, {\bf Holographic Superconductors}, \textit{JHEP} \textbf{11}, 015 (2009) [arXiv:0908.3677 [hep-th]].
\bibitem{horowitzlect} G.T. Horowitz, {\it Introduction to Holographic Superconductors}, Lect.\ Notes Phys.\ \textbf{828}, 313 (2011) [arXiv:1002.1722 [hep-th]].
\bibitem{herzogrev} C.P. Herzog, {\it Lectures on Holographic Superfluidity and Superconductivity}, J.\ Phys.\ A \textbf{42}, 343001 (2009) [arXiv:0904.1975 [hep-th]].

\bibitem{Ling:2017naw} Y. Ling, P. Liu, J.P. Wu and M.H. Wu, {\bf Holographic superconductor on a novel insulator}, \textit{Chin. Phys. C} \textbf{42}, 013106 (2018) [arXiv:1711.07720 [hep-th]].
\bibitem{Ling:2015epa} Y. Ling, P. Liu, C. Niu and J.P. Wu, {\bf Building a doped Mott system by holography}, \textit{Phys. Rev. D} \textbf{92}, 086003 (2015) [arXiv:1507.02514 [hep-th]].
\bibitem{Ling:2014laa} Y. Ling, P. Liu, C. Niu, J.P. Wu and Z.Y. Xian, {\bf Holographic Superconductor on Q-lattice}, \textit{JHEP} \textbf{02}, 059 (2015) [arXiv:1410.6761 [hep-th]].
\bibitem{Zeng:2014uoa} H.B. Zeng and J.P. Wu, {\bf Holographic superconductors from the massive gravity}, \textit{Phys. Rev. D} \textbf{90}, 046001 (2014) [arXiv:1404.5321 [hep-th]].
\bibitem{Liu:2022bdu} Y. Liu, X.J. Wang, J.P. Wu and X. Zhang, {\bf Holographic superfluid with gauge--axion coupling}, \textit{Eur. Phys. J. C} \textbf{83}, 748 (2023) [arXiv:2212.01986 [hep-th]].
\bibitem{Liu:2022bam} Y. Liu, X.J. Wang, J.P. Wu and X. Zhang, {\bf Alternating current conductivity and superconducting properties of a holographic effective model with broken translations}, \textit{Eur. Phys. J. C} \textbf{82}, 478 (2022) [arXiv:2201.06065 [hep-th]].
\bibitem{Liu:2020hhx} Y. Liu, G. Fu, H.L. Li, J.P. Wu and X. Zhang, {\bf Holographic p-wave superconductivity from higher derivative theory}, \textit{Eur. Phys. J. C} \textbf{81}, 568 (2021) [arXiv:2011.07330 [hep-th]].
\bibitem{Li:2019dmm} H.L. Li, G. Fu, Y. Liu, J.P. Wu and X. Zhang, {\bf Higher derivatives driven symmetry breaking in holographic superconductors}, \textit{Eur. Phys. J. C} \textbf{80}, 102 (2020) [arXiv:1910.07694 [hep-th]].
\bibitem{Wu:2017xki} J.P. Wu and P. Liu, {\bf Holographic superconductivity from higher derivative theory}, \textit{Phys. Lett. B} \textbf{774}, 527 (2017) [arXiv:1710.07971 [hep-th]].
\bibitem{Ma:2011zze} D.Z. Ma, Y. Cao and J.P. Wu, {\bf The St\"{u}ckelberg holographic superconductors with Weyl corrections}, \textit{Phys. Lett. B} \textbf{704}, 604 (2011) [arXiv:1201.2486 [hep-th]].
\bibitem{Wu:2010vr} J.P. Wu, Y. Cao, X.M. Kuang and W.J. Li, {\bf The 3+1 holographic superconductor with Weyl corrections}, \textit{Phys. Lett. B} \textbf{697}, 153 (2011) [arXiv:1010.1929 [hep-th]].


\bibitem{siopsis} G. Siopsis and J. Therrien, {\it Analytic calculation of properties of holographic superconductors}, J. High Energ. Phys. {\bf 05}, 013 (2010) [arXiv:1003.4275 [hep-th]].
\bibitem{gregory} R.~Gregory, S.~Kanno and J.~Soda, {\it Holographic Superconductors with Higher Curvature Corrections}, JHEP \textbf{10}, 010 (2009) [arXiv:0907.3203 [hep-th]].
\bibitem{panwang} Q.~Pan and B.~Wang, {\it Analytical study on holographic superconductors with backreactions}, arXiv:1205.3543 [hep-th].
\bibitem{gangopadhyay} S.~Gangopadhyay, {\it Analytic study of properties of holographic superconductors away from the probe limit}, arXiv:1302.1288 [hep-th].
\bibitem{zengsl} H.-B.~Zeng, X.~Gao, Y.~Jiang and H.-S.~Zong, {\it Analytical Computation of Critical Exponents in Several Holographic Superconductors}, arXiv:1012.5564 [hep-th].
\bibitem{Wang:2019fkt}
C.~Wang, D.~Zhang, G.~Fu and J.~P.~Wu,
\textbf{Analytical Study of the Holographic Superconductor from Higher Derivative Theory},
Adv. High Energy Phys. \textbf{2020} (2020), 5902473
[arXiv:1902.07125 [gr-qc]].
\bibitem{kimseo2024} S.~Kim, K.~K.~Kim and Y.~Seo, {\it Phase Diagram from Nonlinear Interaction between Superconducting Order and Density: Toward Data-Based Holographic Superconductor}, arXiv:2410.06523 [hep-th].
\bibitem{brent1973} R. P. Brent, {\it Algorithms for Minimization without Derivatives}, Prentice‑Hall, 1973.
\bibitem{scipy} P. Virtanen et al., {\it SciPy 1.0: Fundamental Algorithms for Scientific Computing in Python}, Nature Methods \textbf{17}, 261–272 (2020).
\bibitem{zhu1997} C. Zhu, R. H. Byrd, P. Lu, and J. Nocedal, {\it Algorithm 778: L‑BFGS‑B: Fortran Subroutines for Large‑Scale Bound‑Constrained Optimization}, ACM Trans. Math. Softw. \textbf{23}, 550–560 (1997).
\bibitem{mckay1979} M. D. McKay, R. J. Beckman, and W. J. Conover, {\it A Comparison of Three Methods for Selecting Values of Input Variables in the Analysis of Output from a Computer Code}, Technometrics \textbf{21}, 239–245 (1979).




\end{thebibliography}
\end{document}